\documentclass[11pt]{article}
\usepackage[margin=1in]{geometry}
\usepackage[final]{microtype}
\usepackage[T1]{fontenc}
\usepackage{amsmath,amssymb,amsthm}
\usepackage[hidelinks]{hyperref}

\newtheorem{theorem}{Theorem}
\newtheorem{lemma}{Lemma}

\newtheorem{observation}{Observation}
\theoremstyle{definition}

\theoremstyle{remark}

\newcommand{\E}{\mathbb{E}}
\newcommand{\OPT}{\operatorname{OPT}}

\title{An Improved Lower Bound for Randomized Online Scheduling}
\author{Jianan Lin\\[0.75em]
Rensselaer Polytechnic Institute}
\date{}

\begin{document}
\maketitle

\begin{abstract}
We improve the asymptotic lower bound for randomized online scheduling on identical machines from $\frac{e}{e-1}\approx 1.58198$ [Chen, van Vliet, and Woeginger, IPL 1994; Sgall, IPL 1997] to $\frac{e}{e-1}+\frac{1}{715}\approx 1.58338$.
Although the improvement is small, it is the first in 32 years.
\end{abstract}

\section{Introduction}\label{sec:introduction}

In online scheduling on $m$ identical machines, jobs arrive one by one.
When a job arrives, the algorithm learns its processing time and must assign it to a machine immediately and irrevocably.
The goal is to minimize the makespan, which is the maximum load of a machine.
An online algorithm is $R$-competitive if, on every input, its expected makespan is at most $R$ times the optimal makespan of that input.

Graham~\cite{graham1966bounds} showed that list scheduling, which assigns each incoming job to a least-loaded machine, is $\big(2-\frac{1}{m}\big)$-competitive.
Bartal et al.~\cite{bartal1995ancient} gave the first deterministic algorithm with a competitive ratio bounded away from $2$ as $m$ grows, achieving $1.986$ for sufficiently large $m$.
Albers~\cite{albers1999better} brought the upper bound down to $1.923$, and Fleischer and Wahl~\cite{fleischer2000revisited} obtained an asymptotic upper bound below $1.9201$.
Rudin~\cite{rudin2001improved} proved a deterministic lower bound of $1.88$, which Xiao and Tan~\cite{xiao2026improved} recently improved to $1.887111$ with a construction on $8192$ machines.

Randomization helps even on two machines, where Bartal et al.~\cite{bartal1995ancient} obtained the optimal ratio of $\frac{4}{3}$, compared with $\frac{3}{2}$ for deterministic algorithms.
For general $m$, Albers~\cite{albers2002randomized} gave a randomized algorithm with an asymptotic competitive ratio of $1.916$.
On the lower-bound side, Chen, van Vliet, and Woeginger~\cite{chen1994} and Sgall~\cite{sgall1997} proved a randomized lower bound of $\frac{1}{1-\left(1-\frac{1}{m}\right)^m}$, which tends to $\frac{e}{e-1}\approx 1.58198$ as $m\to\infty$.
Tich\'y~\cite{tichy2004} improved this bound for three machines, but the asymptotic lower bound of $\frac{e}{e-1}$ has stood since 1994.

\paragraph{Our Results.}
Let $R_m^*$ denote the best competitive ratio achievable by randomized online algorithms on $m$ machines.
We prove that
\[
    R_m^*\ge \frac{e}{e-1}+\frac{1}{715}\qquad\text{for every } m\ge 10^8.
\]
The classical bound $\frac{1}{1-(1-\frac1m)^m}$ is smaller than $\frac{e}{e-1}$ for every $m$, so our bound exceeds it by more than $\frac{1}{715}$, and the asymptotic lower bound rises from $\frac{e}{e-1}$ to $\frac{e}{e-1}+\frac{1}{715}$.

The classical lower bound uses a single sequence of geometrically increasing jobs that may stop after any of its last $m$ jobs.
Since the algorithm does not know where the sequence stops, it must keep its makespan small at every possible stopping point.
We keep this sequence as a common prefix, but let some inputs continue with one or two batches of larger jobs.
An algorithm can keep the makespan low after the first batch by spreading it over many machines, but then few machines remain free for the second batch.
This tradeoff alone gives a constant improvement over $\frac{e}{e-1}$ (Section~\ref{sec:warmup}).
For the improvement of $\frac{1}{715}$ (Section~\ref{sec:main-lower}), we also change the prefix: a block of equal-size jobs in its middle lowers the sum of the offline optima of the other stopping points.

\paragraph{Additional Related Work.}
For every fixed $m$, Megow and Wiese~\cite{megow2013schemes} gave a scheme that approximates the optimal deterministic competitive ratio to any desired precision and constructs an algorithm whose ratio is arbitrarily close to it.
For four machines, Janke~\cite{janke2026lower} recently raised the deterministic lower bound from $\sqrt{3}$ to $\frac{14350}{8283}$, with an adversary that chooses the next jobs based on the algorithm's earlier assignments.
Albers and Janke~\cite{albers2021randomorder} studied the random-order model, in which an adversary chooses the set of jobs and the jobs arrive in a uniformly random order.
In online bin stretching, a semi-online variant, the algorithm knows the optimal makespan in advance; for this variant, B{\"o}hm et al.~\cite{bohm2024knownmakespan} gave a $\frac{139}{93}$-competitive algorithm for sufficiently many machines.

\section{Preliminaries}\label{sec:preliminaries}

There are $m\ge2$ identical machines, indexed by $[m]=\{1,\ldots,m\}$.
An input $I$ is a finite sequence of jobs, each with a positive processing time, which we also call its size.
The jobs are revealed one at a time, and an online algorithm must assign each job to a machine before the next one is revealed, without knowing how many jobs remain.
For an algorithm $A$, let $C_A(I)$ be its makespan on $I$, that is, the maximum total size of the jobs assigned to a machine.
Let $\OPT(I)$ be the minimum makespan over all assignments of $I$; it is at least the size of the largest job in $I$.

A randomized online algorithm $A$ is $R$-competitive if $\E[C_A(I)]\le R\OPT(I)$ for every input $I$, where the expectation is over the random choices of $A$.
Thus $R_m^*$ is the infimum of $R$ over all $R$-competitive randomized online algorithms on $m$ machines.

We may assume that a randomized online algorithm draws a random string before the first job arrives, so that fixing this string gives a deterministic online algorithm.
A deterministic online algorithm cannot see future jobs, so it assigns the jobs of a common prefix in the same way on every input that starts with this prefix.
All our lower bounds rest on the following standard averaging argument.

\begin{observation}\label{obs:averaging}
Let $I^1,\ldots,I^k$ be inputs and $X>0$.
If every deterministic online algorithm $A$ satisfies $\sum_{j=1}^k C_A(I^j)\ge X$, then
\[
    R_m^*\ge\frac{X}{\sum_{j=1}^k\OPT(I^j)}.
\]
\end{observation}

\begin{proof}
Let $A$ be an $R$-competitive randomized online algorithm.
For every fixed random string, $A$ is a deterministic online algorithm, so $\sum_j C_A(I^j)\ge X$ holds for every realization.
Taking expectations and applying the competitiveness of $A$ to each input gives $X\le\sum_j\E[C_A(I^j)]\le R\sum_j\OPT(I^j)$.
\end{proof}

Throughout the paper, we write
\[
    q=1-\frac{1}{m},\qquad \sigma_m=\frac{1}{1-q^m}.
\]
The classical lower bound is $R_m^*\ge\sigma_m$~\cite{chen1994,sgall1997}.
Since $q^m$ increases to $e^{-1}$, the sequence $\sigma_m$ increases to $\frac{e}{e-1}$.
We write $\log$ for the natural logarithm.

\section{Warmup: A Strict Improvement}\label{sec:warmup}

We first show that the classical construction can be strengthened without changing its prefix.
In the classical argument every input is a prefix of one common sequence; we keep this sequence but replace some of its stopping points by inputs that continue with one or two batches of larger jobs.

\begin{theorem}\label{thm:warmup}
There exists a constant $\delta>0$ such that
\[
    \liminf_{m\to\infty}R_m^*\ge\frac{e}{e-1}+\delta.
\]
\end{theorem}

The common prefix consists of small jobs with total size $mq^m$, followed by $m$ larger jobs of sizes $p_i=q^{m-i}$, $i\in[m]$; the index $i$ counts only these larger jobs.
Let $I_i$ be the input that ends with the job of size $p_i$.
Since $mq+1=m$, induction on $i$ shows that $I_i$ has total size $mp_i$; in particular, the prefix has total size $m$.

Let $h=\bigl\lfloor\frac{m}{2}\bigr\rfloor$.
For each integer $1\le r<h$, set $k_r=m-h+r$ and $u_r=\frac{m}{h-r}$.
The input $J_r$ appends a first batch of $k_r$ jobs of size $u_r$ to the prefix, and $J'_r$ appends a second batch of $h$ jobs of size $2u_r$ to $J_r$.
All small jobs have size $\frac1N$, where $N$ is a common denominator of $q^m$, the $p_i$, and the $u_r$; this makes exact offline packings possible.

The offline optima of $I_i$, $J_r$, and $J'_r$ are $p_i$, $u_r$, and $2u_r$, respectively.
Each value is the size of the largest job in the input, so it is a lower bound.
For $I_i$, put the jobs of sizes $p_1,\ldots,p_i$ on separate machines of capacity $p_i$, and fill the remaining capacity, which is at least $mq^m$, with small jobs.
For $J_r$, put the first batch on $k_r$ separate machines of capacity $u_r$.
The other $h-r$ machines then have total capacity $(h-r)u_r=m$, exactly the total size of the prefix.
For $J'_r$, use capacity $2u_r$: put the second batch on $h$ machines and the first batch on the other $m-h$ machines, at most two per machine, which is possible since $k_r\le2(m-h)$.
The remaining capacity is $(m-h-r)u_r\ge(h-r)u_r=m$.
In both cases we pack the prefix in reverse order, starting with $p_m=1$.
Before the job of size $p_i$ is placed, the remaining capacity is at least $mp_i$, so some machine can take it, and afterwards at least $(m-1)p_i=mqp_i$ remains for the next job.
The small jobs fit at the end because all remaining capacities are multiples of $\frac1N$ and sum to at least $mq^m$.

Now consider an online algorithm facing $J'_r$.
If it spreads the first batch over $k_r$ distinct machines, only $m-k_r=h-r<h$ machines are still free when the $h$ jobs of the second batch arrive.
If it puts two first-batch jobs on one machine, its makespan already reaches $2u_r$ after the first batch.
The following lemma turns this tradeoff into a bound on the sum of the two makespans.

\begin{lemma}\label{lem:main-two-batches}
Suppose the machine loads before the two batches are $\ell_1\le\cdots\le\ell_m$ and sum to $m$.
Fix $1\le r<h$, and let $C_0$ and $C_1$ be the makespans after the first and second batches of $J'_r$.
Then
\[
    C_0+C_1\ge 4u_r+\ell_r+\ell_{h+r}.
\]
\end{lemma}

\begin{proof}
Write $k=k_r$ and $u=u_r$.
The $m-k+1=h-r+1$ largest initial loads are each at least $\ell_k$ and sum to at most $m$, so $\ell_k\le\frac{m}{h-r+1}<u$.
Moreover, $h+r\le m-h+r=k$, so $\ell_r\le\ell_{h+r}\le\ell_k<u$.
If the first batch uses $k$ distinct machines, one of them has initial load at least $\ell_k$, so $C_0\ge u+\ell_k$; otherwise some machine receives two first-batch jobs and $C_0\ge 2u>u+\ell_k$.
Hence $C_0\ge u+\ell_k$ in every case.

If some machine receives three first-batch jobs, then $C_0,C_1\ge 3u$, and $C_0+C_1\ge 6u>4u+\ell_r+\ell_{h+r}$.
If some machine receives two second-batch jobs, or two first-batch jobs and a second-batch job, then $C_1\ge 4u$, and together with $C_0\ge u+\ell_k$ we get $C_0+C_1\ge 5u+\ell_k>4u+\ell_r+\ell_{h+r}$.

In the remaining cases, every machine receives at most two first-batch jobs and at most one second-batch job, and no machine receives both two first-batch jobs and a second-batch job.
Let $d$ be the number of machines with two first-batch jobs, and let $o$ be the number of machines with one job from each batch.
The two batches occupy $k-d$ and $h$ machines, respectively, and share $o$ of them, so $k-d+h-o\le m$, that is, $d+o\ge r$.

\begin{itemize}
    \item If $d=0$, then at least $r$ machines receive a job from each batch, so $C_1\ge 3u+\ell_r$. Adding $C_0\ge u+\ell_k\ge u+\ell_{h+r}$ proves the bound.
    \item If $o=0$, the $d\ge r$ machines with two first-batch jobs and the $h$ machines with a second-batch job form two disjoint groups. Each group has at least $r$ machines, and together they have at least $h+r$, so the smaller of their maximum initial loads is at least $\ell_r$ and the larger is at least $\ell_{h+r}$. The machines of the first group gain $2u$ in the first batch and those of the second group gain $2u$ in the second batch, so $C_0+C_1\ge 4u+\ell_r+\ell_{h+r}$.
    \item If $d,o>0$, let $M_d$ and $M_o$ be the maximum initial loads among the $d$ machines with two first-batch jobs and among the $o$ machines with one job from each batch. Then $C_0\ge 2u+M_d$ and $C_1\ge 3u+M_o$. The two groups are disjoint and together contain $d+o\ge r$ machines, so $\max\{M_d,M_o\}\ge\ell_r$. As loads are nonnegative, $C_0+C_1\ge 5u+\ell_r>4u+\ell_r+\ell_{h+r}$. \qedhere
\end{itemize}
\end{proof}

\begin{proof}[Proof of Theorem~\ref{thm:warmup}]
Fix a deterministic online algorithm, write $C(I)$ for its makespan on input $I$, and let $\ell_1\le\cdots\le\ell_m$ be its loads at the end of the prefix.
Only $m-i$ jobs of the prefix arrive after $I_i$, so at least $i$ machines receive no further load, and hence $C(I_i)\ge\ell_i$.
Summing over all $i$ gives $\sum_i C(I_i)\ge m$, while $\sum_i\OPT(I_i)=\sum_i p_i=m(1-q^m)$; by Observation~\ref{obs:averaging}, this is the classical bound $\sigma_m$.

To improve it, choose an integer $1\le K<h$ and, for each $r\in[K]$, replace $I_r$ and $I_{h+r}$ by $J_r$ and $J'_r$.
Let $S=\{1,\ldots,K\}\cup\{h+1,\ldots,h+K\}$ be the indices of the removed inputs.
The algorithm treats $J_r$ and $J'_r$ identically until the end of the first batch, so Lemma~\ref{lem:main-two-batches} gives $C(J_r)+C(J'_r)\ge 4u_r+\ell_r+\ell_{h+r}$.
Together with $C(I_i)\ge\ell_i$ for $i\notin S$, this yields
\begin{equation}\label{eq:two-batch-sum}
    \sum_{i\notin S}C(I_i)+\sum_{r=1}^K\bigl(C(J_r)+C(J'_r)\bigr)\ge m+4\sum_{r=1}^K u_r.
\end{equation}
Each replacement changes the sum of offline optima by $3u_r-p_r-p_{h+r}$.
Writing
\[
    H_m=\frac1m\sum_{r=1}^K u_r=\sum_{r=1}^K\frac{1}{h-r},\qquad
    T_m=\frac1m\sum_{r=1}^K(p_r+p_{h+r}),
\]
the right-hand side of~\eqref{eq:two-batch-sum} is $m(1+4H_m)$ and the offline optima of the selected inputs sum to $m(1-q^m+3H_m-T_m)$.
Observation~\ref{obs:averaging} therefore gives
\[
    R_m^*\ge\frac{1+4H_m}{1-q^m+3H_m-T_m}.
\]

Now let $K=\lfloor xm\rfloor$ for a fixed $x\in\bigl(0,\frac12\bigr)$.
Since $H_m$ is a Riemann sum and $T_m$ is a sum of two geometric progressions, as $m\to\infty$ we have
\[
    \begin{aligned}
    H_m&\to H(x)=\int_0^x\frac{dt}{\frac{1}{2}-t}=-\log(1-2x),\\
    T_m&=(q^{m-K}+q^{m-h-K})(1-q^K)\to T(x)=\bigl(e^{-1}+e^{-\frac{1}{2}}\bigr)(e^x-1).
    \end{aligned}
\]
The limit of the lower bound is $\frac{1+4H(x)}{1-e^{-1}+3H(x)-T(x)}$, whose denominator is positive for small $x$.
Multiplying out shows that it exceeds $\frac{1}{1-e^{-1}}=\frac{e}{e-1}$ if and only if
\[
    T(x)>\Bigl(\frac{4}{e}-1\Bigr)H(x).
\]
Both sides vanish at $x=0$, and the derivative of their difference at $x=0$ is
\[
    e^{-1}+e^{-\frac{1}{2}}-2\Bigl(\frac{4}{e}-1\Bigr)
    =2-\frac{7}{e}+e^{-\frac{1}{2}}
    >2-\frac{70}{27}+\frac{3}{5}=\frac{1}{135},
\]
where we used $e>\frac{27}{10}$ and $\sqrt{e}<\frac{5}{3}$.
Hence the inequality holds for all sufficiently small $x>0$, and fixing such an $x$ gives the claimed $\delta$.
\end{proof}

\section{The Main Lower Bound}\label{sec:main-lower}

The improvement in Section~\ref{sec:warmup} comes entirely from the inputs $J_r$ and $J'_r$; the remaining stopping points are still those of the classical construction.
We now change the prefix so that the remaining stopping points have a smaller sum of offline optima.
The idea is to make $D\ge K$ consecutive larger jobs, starting right after the $h$-th one, all equal to some size $A$.
This block contains the jobs at positions $h+1,\ldots,h+K$, whose stopping points are replaced anyway.
Equal-size jobs carry more of the total size than the geometric jobs they replace, so the jobs before the block become smaller, and so do the optima of $I_{K+1},\ldots,I_h$, at the cost of larger optima for the stopping points $I_{h+K+1},\ldots,I_{h+D}$ inside the block.
Inequality~\eqref{eq:two-batch-sum} does not depend on the sizes of the prefix jobs, so it remains valid.

\begin{theorem}\label{thm:main-lower}
For every $m\ge10^8$, we have
\[
    R_m^*\ge\frac{e}{e-1}+\frac{1}{715}.
\]
In particular, $R_m^*>\sigma_m+\frac{1}{715}$ for every $m\ge 10^8$.
\end{theorem}

We use $q$, $h$, $k_r$, and $u_r$ as in Section~\ref{sec:warmup}.
Let $K$ and $D$ be integers with $1\le K<h$ and $K\le D\le h$; their values are fixed in the proof of Theorem~\ref{thm:main-lower}.
Define
\[
    A=q^{m-h-D},\qquad a=\left(1-\frac{D}{m}\right)A,\qquad b=aq^h,
\]
and let $N=m^m\operatorname{lcm}(1,\ldots,h)$.
The prefix starts with $mbN$ small jobs of size $\frac{1}{N}$, of total size $mb$, followed by $m$ larger jobs of sizes
\[
    p_i=
    \begin{cases}
        aq^{h-i}, & 1\le i\le h,\\
        A, & h<i\le h+D,\\
        q^{m-i}, & h+D<i\le m.
    \end{cases}
\]
The sizes grow geometrically up to $a$, stay at $A$ for $D$ jobs, and then grow geometrically up to $1$.
With $D=0$, this would be exactly the prefix of Section~\ref{sec:warmup}.
The factor $m^m$ clears the denominators of the $p_i$ and of $b$, and $\operatorname{lcm}(1,\ldots,h)$ clears those of the $u_r$.
Hence all job sizes, as well as the capacities $u_r$ and $2u_r$ used below, are multiples of $\frac1N$, and $mbN$ is an integer.
The inputs $I_i$, $J_r$, and $J'_r$ are defined as in Section~\ref{sec:warmup}, now for $r\in[K]$.

\begin{lemma}\label{lem:main-opt}
The prefix has total size $m$.
Moreover, $\OPT(I_i)=p_i$ for $i\in[m]$, and $\OPT(J_r)=u_r$ and $\OPT(J'_r)=2u_r$ for $r\in[K]$.
\end{lemma}

\begin{proof}
For $i\le h$, the total size through the job of size $p_i$ is $mp_i$: the small jobs have total size $mb=mqp_1$, and if the total before $p_i$ is $mqp_i$, adding $p_i$ gives $mp_i$.
At $i=h$, the total is $ma=(m-D)A$.
The $D$ jobs of size $A$ raise it to $mA$, and the same calculation gives total $mp_i$ for every $i>h+D$, ending at $mp_m=m$.
Hence $I_i$ has total size at most $mp_i$ for every $i$.
Putting the jobs of sizes $p_1,\ldots,p_i$ on separate machines of capacity $p_i$ and placing the small jobs in the remaining capacity gives makespan $p_i$; the job of size $p_i$ shows that this is optimal.

For $J_r$, putting each first-batch job on its own machine of capacity $u_r$ leaves total capacity $(m-k_r)u_r=m$.
For $J'_r$, with capacity $2u_r$, putting the second batch on $h$ separate machines and the first batch on the other $m-h$ machines, at most two per machine, leaves total capacity $(m-h-r)u_r\ge m$.
In both cases at least $h$ machines are full, so at most $m-h$ machines have positive remaining capacity.

We pack the prefix into this capacity in reverse order.
For the last geometric part ($i>h+D$), the argument of Section~\ref{sec:warmup} applies: before a job of size $x$ is placed, at least $mx$ capacity remains, and afterwards at least $mqx$ remains.
If this part is nonempty, its smallest job has size $\frac{A}{q}$, so at least $mA$ capacity remains after it; if it is empty, then $A=1$ and the available capacity is already at least $m=mA$.
If $s_j$ denotes the remaining capacities of the at most $m-h$ machines that are not full, then the number of jobs of size $A$ that fit is at least
\[
    \sum_j\left\lfloor\frac{s_j}{A}\right\rfloor
    \ge \frac{\sum_j s_j}{A}-(m-h)
    \ge m-(m-h)
    = h \ge D.
\]
After these $D$ jobs, at least $(m-D)A=ma$ capacity remains, so the same reverse-order argument places $p_h,\ldots,p_1$.
At least $mb$ capacity then remains, in multiples of $\frac1N$, for the small jobs.
This gives schedules of makespan $u_r$ and $2u_r$, which are optimal because of the largest jobs.
\end{proof}

\begin{proof}[Proof of Theorem~\ref{thm:main-lower}]
For each $r\in[K]$, replace $I_r$ and $I_{h+r}$ by $J_r$ and $J'_r$, and let $S=\{1,\ldots,K\}\cup\{h+1,\ldots,h+K\}$ as before.
The proof of~\eqref{eq:two-batch-sum} used only that the prefix has total size $m$ and that $m-i$ of its larger jobs arrive after $I_i$, so~\eqref{eq:two-batch-sum} still holds by Lemma~\ref{lem:main-opt}.

By Lemma~\ref{lem:main-opt}, the remaining prefix inputs have offline optima summing to $\sum_{i\notin S}p_i$.
The larger jobs have total size $\sum_{i=1}^m p_i=m(1-b)$.
The removed inputs $I_1,\ldots,I_K$ have optima summing to $\sum_{i\le K}p_i=m(p_K-b)$, and each of $I_{h+1},\ldots,I_{h+K}$ has optimum $A$.
Since $p_K=aq^{h-K}=\bigl(1-\frac{D}{m}\bigr)q^{m-D-K}$ and $A=q^{m-h-D}$, we obtain
\[
    \sum_{i\notin S}p_i
    = m(1-b)-m(p_K-b)-KA
    = mQ_m,
    \qquad\text{where }
    Q_m=1-\frac{K}{m}q^{m-h-D}-\left(1-\frac{D}{m}\right)q^{m-D-K}.
\]
With $H_m$ as in Section~\ref{sec:warmup}, the offline optima of all selected inputs sum to $m(Q_m+3H_m)$, and Observation~\ref{obs:averaging} gives
\[
    R_m^*\ge \Phi_m:=\frac{1+4H_m}{Q_m+3H_m}.
\]

We take $K=\left\lfloor\frac{19m}{500}\right\rfloor$ and $D=\left\lfloor\frac{3m}{50}\right\rfloor$, which satisfy the constraints above.
As $m\to\infty$, the quantities $H_m$, $Q_m$, and $\Phi_m$ tend to
\[
    H=\log\frac{250}{231},\qquad
    Q=1-\frac{19}{500}e^{-\frac{11}{25}}-\frac{47}{50}e^{-\frac{451}{500}},\qquad
    \Phi_\infty=\frac{1+4H}{Q+3H}.
\]
Numerically, $H\approx0.0790432$, $Q\approx0.5941147$, and $\Phi_\infty\approx1.5833766$, which exceeds $\frac{e}{e-1}+\frac{1}{715}\approx1.5833753$ by about $1.27\times10^{-6}$.
The rest of the proof makes this rigorous for every $m\ge10^8$: we bound $|H_m-H|$ and $|Q_m-Q|$, certify the numerical values, and then control $|\Phi_m-\Phi_\infty|$.

The integrand of $\int_0^K\frac{dx}{h-x}=\log\frac{h}{h-K}$ is increasing, and $H_m$ is its right-endpoint Riemann sum, so
\[
    0\le H_m-\log\frac{h}{h-K}\le \frac{1}{h-K}-\frac{1}{h}<\frac{0.2}{m}.
\]
On the rectangle with opposite corners $\bigl(\frac{h}{m},\frac{K}{m}\bigr)$ and $\bigl(\frac{1}{2},\frac{19}{500}\bigr)$, the partial derivatives of $\log\frac{t}{t-w}$ with respect to $t$ and $w$ are below $0.2$ and $2.2$ in absolute value, and the two coordinates differ by at most $\frac{1}{2m}$ and $\frac{1}{m}$.
By the mean value theorem, $\bigl|\log\frac{h}{h-K}-H\bigr|<\frac{2.3}{m}$, and hence $|H_m-H|<\frac{4}{m}$.

For $Q_m$, write $q^k=e^{-k(-\log q)}$.
The series $-\log(1-x)=\sum_{j\ge1}\frac{x^j}{j}$ gives
\[
    0\le -\log q-\frac{1}{m}\le\frac{1}{2m(m-1)},
\]
so for $k\le m$, the quantity $k(-\log q)$ differs from $\frac{k}{m}$ by at most $\frac{1}{2(m-1)}$.
Because of the rounding in $h$, $K$, and $D$, the ratios $\frac{m-h-D}{m}$ and $\frac{m-D-K}{m}$ differ from $\frac{11}{25}$ and $\frac{451}{500}$ by at most $\frac{1.5}{m}$ and $\frac{2}{m}$.
Hence $(m-h-D)(-\log q)$ and $(m-D-K)(-\log q)$ differ from $\frac{11}{25}$ and $\frac{451}{500}$ by at most $\frac{2.1}{m}$ and $\frac{2.6}{m}$.
Since $x\mapsto e^{-x}$ is $1$-Lipschitz on $[0,\infty)$, the powers $q^{m-h-D}$ and $q^{m-D-K}$ are within $\frac{2.1}{m}$ and $\frac{2.6}{m}$ of $e^{-\frac{11}{25}}$ and $e^{-\frac{451}{500}}$.
The coefficients $\frac{K}{m}$ and $1-\frac{D}{m}$ are within $\frac1m$ of $\frac{19}{500}$ and $\frac{47}{50}$, and all four factors lie in $[0,1]$, so $|Q_m-Q|<\frac{5}{m}$.

To certify the numerical values, we need three inequalities: $Q+3H>0.81$, $\Phi_\infty<1.6$, and
\[
    \Phi_\infty-\frac{e}{e-1}-\frac{1}{715}>\frac{12}{10^7}.
\]
For $x\in\bigl\{\frac{11}{25},\frac{451}{500},1\bigr\}$, the Taylor polynomials of $e^{-x}$ of degrees $13$ and $12$ are rational lower and upper bounds, respectively.
For $H$, let $z=\frac{19}{481}$, so that $\frac{250}{231}=\frac{1+z}{1-z}$.
The series $\log\frac{1+z}{1-z}=2\sum_{j\ge0}\frac{z^{2j+1}}{2j+1}$ then gives
\[
    H_0:=2\sum_{j=0}^5\frac{z^{2j+1}}{2j+1}
    \le H
    \le H_0+\frac{2z^{13}}{13(1-z^2)}.
\]
The three inequalities follow from these rational bounds by exact arithmetic.

Finally, for $m\ge10^8$ we have $|Q_m-Q|+3|H_m-H|<\frac{17}{m}<0.01$, so $Q_m+3H_m>0.8$.
Since
\[
    \Phi_m-\Phi_\infty
    = \frac{4(H_m-H)-\Phi_\infty\bigl((Q_m-Q)+3(H_m-H)\bigr)}{Q_m+3H_m},
\]
we get
\[
    |\Phi_m-\Phi_\infty|
    \le \frac{4|H_m-H|+\Phi_\infty\bigl(|Q_m-Q|+3|H_m-H|\bigr)}{Q_m+3H_m}
    <\frac{16+1.6\cdot17}{0.8\,m}=\frac{54}{m}.
\]
For $m\ge10^8$, this is less than $\frac{12}{10^7}$, so $R_m^*\ge\Phi_m>\frac{e}{e-1}+\frac{1}{715}$.
The final claim follows from $\sigma_m<\frac{e}{e-1}$.
\end{proof}

\section{Conclusion}\label{sec:conclusion}

We showed that the randomized competitive ratio of online scheduling on $m$ identical machines is at least $\frac{e}{e-1}+\frac{1}{715}$ for every $m\ge10^8$, improving the bound of Chen, van Vliet, and Woeginger~\cite{chen1994} and Sgall~\cite{sgall1997}.
The bound $\frac{e}{e-1}$ is elegant, but it is not the final answer.
The improvement comes from appending two batches of large jobs to the classical prefix: an algorithm that keeps its makespan low after the first batch pays for it after the second.
A block of equal-size jobs inside the prefix then lowers the sum of the offline optima of the other stopping points.
The gap to the best known upper bound of $1.916$~\cite{albers2002randomized} remains large.
Narrowing it, and improving the lower bound for small numbers of machines, are natural next steps.

\paragraph{AI Use Statement.}
The author used GPT Astra-6 to explore proofs of the improved lower bound and to better understand the proof technique.
The author used Codex, Gemini 4 Neon, and Claude Opus to polish the writing.

\bibliographystyle{plain}
\bibliography{mybibfile}

\end{document}